# Position and momentum mapping of vibrations in graphene nanostructures in the electron microscope

Ryosuke Senga and Kazu Suenaga*

*[1]Nanomaterials Research Institute, National Institute of Advanced Industrial Science and Technology (AIST), Tsukuba 305-8565, Japan.*

Paolo Barone

*[2]SPIN-CNR, c/o Università G. D'Annunzio, I-66100 Chieti, Italy*

Shigeyuki Morishita

*[3]JEOL Ltd., 3-1-2 Musashino, Akishima, Tokyo 196-8558, Japan*

Francesco Mauri

*[4]Dipartimento di Fisica, Università di Roma La Sapienza, Piazzale Aldo Moro 5, I-00185, Roma, Italy*

*[5]Graphene Labs, Fondazione Istituto Italiano di Tecnologia, Via Morego, I-16163 Genova, Italy*

Thomas Pichler

*[6]Faculty of Physics, University of Vienna, Strudlhofgasse 4, A-1090 Vienna, Austria*

Propagating atomic vibrational waves, phonons, rule important thermal, mechanical, optoelectronic and transport characteristics of materials. Thus the knowledge of phonon dispersion, namely the dependence of vibrational energy on momentum is a key ingredient to understand and optimize the material's behavior. However, despite its scientific importance in the last decade, the phonon dispersion of a freestanding monolayer of two dimensional (2D) materials such as graphene and its local variations has still remained elusive because of experimental limitations of vibrational spectroscopy. Even though electron energy loss spectroscopy (EELS) in transmission has recently been shown to probe the local vibrational charge responses[1-4], these studies are yet limited to polar materials like boron nitride or oxides[1-4], in which huge signals induced by strong dipole moments are present. On the other hand, measurements on graphene performed by inelastic x-ray (neutron) scattering spectroscopy[5-7] or EELS in reflection[8, 9] do not have any spatial resolution and require large microcrystals. Here we provide a new pathway to determine the phonon dispersions down to the scale of an individual freestanding graphene monolayer by mapping the distinct vibration modes for a large momentum transfer. The measured scattering intensities are accurately reproduced and interpreted with density functional perturbation theory (DFPT)[10]. Additionally, a nanometre-scale mapping of selected *momentum (q)* resolved vibration modes using graphene nanoribbon structures has enabled us to spatially disentangle bulk, edge and surface vibrations.

Our starting point was the measurement of the phonon dispersion from both freestanding graphite and hexagonal boron nitride (h-BN) flakes by obtaining the momentum resolved EEL spectra along the ΓMΓM and ΓKMKΓ directions in the momentum space. The flakes had approximately 10~20 layers and were set perpendicularly to the electron beam (Fig. 1a). A series of twenty EEL spectra were recorded at every 0.25 Å$^{-1}$ along ΓMΓM directions for both graphite and h-BN flakes (left panels in Fig. 1b and 1c, respectively). The raw EEL spectra are shown in Extended Data Fig. 1. The quasi elastic line (zero-loss peak) has been subtracted from each spectrum. In Figs. 1d and 1e (top panels), the intensity colour maps are presented together with the simulated phonon dispersion curves (solid lines) consisting of the transverse, longitudinal and out of plane optical (acoustic) (TO (TA), (LO (LA) and ZO (ZA)) modes, as well as the phonon energy (open triangles) as derived from a Voigtian line-shape analysis on each EEL spectrum (see Extended Data Fig. 2). Since these experimental phonon energies well trace the simulated phonon dispersion curves, the active vibration modes are directly identified and marked (LA, LO,...) in the coloured maps. Spectra collected along the ΓKMKΓ direction are also shown in Extended Data Fig. 3 and 4.

Comparing the vibrational spectra of apolar (graphite) and polar (h-BN) materials, the most significant differences are seen in the first Brillouin zone: the strong peaks arising from the scattering of the LO mode in h-BN (around 170-200 meV) are not observed in graphite. This confirms that apolar materials hardly display IR-active modes because a strong dipole moment is required to be detectable by EELS in transmission or by IR spectroscopy at optical limit ($q$=0 Å$^{-1}$). On the other hand, the LA mode of graphite is already active in the first Brillouin zone and experimentally visible as soon as $q \geq 0.75$ A$^{-1}$, as in h-BN. Then both LA and LO modes become active in the second Brillouin zone, where graphite and h-BN show very similar trends. These facts strongly suggest that the selection rule in outer Brillouin zones is independent on the IR polarizability of materials.

To understand the observed EEL intensities, we express the Stokes cross section of the scattering electrons[11–13] as:

$$\frac{d^2\sigma}{d\Omega d\omega}(\boldsymbol{q}, \omega) = \frac{4}{a_0^2}\frac{1}{q^2}\sum_\nu \frac{1+n_{\boldsymbol{q}\nu}}{\omega_{\boldsymbol{q}\nu}}\left|\sum_I \frac{1}{\sqrt{M_I}}\boldsymbol{Z}_I(\boldsymbol{q})\cdot\boldsymbol{e}_{\boldsymbol{q},\nu}^I e^{-i\boldsymbol{q}\cdot\boldsymbol{\tau}_I}\right|^2 \delta(\omega - \omega_{\boldsymbol{q}\nu}), \qquad (1)$$

where $a_0$ is the Bohr radius, $\boldsymbol{q}$ is the momentum transfer, $I$ and $\nu$ label the atom in the unit cell and the phonon branch, $\omega_{\boldsymbol{q}\nu}$, $n_{\boldsymbol{q}\nu}$ and $\boldsymbol{e}_{\boldsymbol{q},\nu}^I$ are the phonon frequency, occupation and polarization, $M_I$ and $\boldsymbol{\tau}_I$ the atomic mass and position. The $\alpha$ component of the effective charge Cartesian vector $\boldsymbol{Z}_I(\boldsymbol{q})$, accounting for the charge modulation induced by the lattice vibrations, is:

$$Z_{I,\alpha}(\boldsymbol{q}) = -i\frac{V}{q}\Delta n_{I,\alpha}(\boldsymbol{q})e^{i\boldsymbol{q}\cdot\boldsymbol{\tau}_I} + \frac{q_\alpha}{q}\ Z_I^{ion}, \quad (2)$$

where $V$ is the unit-cell volume, $Z_I^{ion} = Z_I - Z_I^{core}$, $Z_I$ and $Z_I^{core}$ are the ionic, nuclear and core-electron (the 1s$^2$ states in the present case, considered as rigid point charges as far as their spatial extension is much smaller than $1/q$) charges. $\Delta n_{I,\alpha}(\boldsymbol{q})$ is the $\boldsymbol{q}$ component of the Fourier transform of the valence-electron density response to the displacement in the $\alpha$ direction of the $I$ ion, that we compute in DFPT, Eq. (12) of Ref. [10], evaluated at the reciprocal lattice vector $\boldsymbol{G} = \boldsymbol{0}$. In this way we fully take into account the effect of valence-electron screening beyond the spherical rigid-ion approximation used in Ref. [13].

The simulated EEL spectra for graphite and h-BN, Figs. 1b and 1c (right panels), show an excellent agreement with the experimental ones. Theory only misses the weak experimental intensity of the ZO mode, which could be activated by the presence of a non-zero out-of-plane momentum provided, e.g., by a sample tilt (see Extended Data Fig. 5). An excellent agreement is also found for the momentum dependence of the total EEL intensity, Fig. 2a, which exhibits a very different behaviour at small $q$ in graphite and h-BN. To rationalize this finding, we analyse the momentum dependence of $\boldsymbol{Z}_I(\boldsymbol{q})$, Fig. 2b. In the long-wavelength limit $q \to 0$, the ionic charge $Z_I^{ion}$ is well screened by the valence density. The screening is perfect in metals (as graphite), so that $\lim_{q\to 0} Z_{I,\alpha}(\boldsymbol{q}) = 0$, implying a vanishing of the EEL signal. In insulators, the screening is partial and, using the expression derived by Vogl in the context of the Frölich interaction[14], one finds that

$$\lim_{q\to 0} Z_{I,\alpha}(\boldsymbol{q}) = q \frac{\sum_\beta Z_{I,\beta\alpha}^B q_\beta}{\sum_{\beta,\gamma} q_\beta \epsilon_{\beta\gamma}^\infty q_\gamma}, \qquad (3)$$

where $Z_{I,\beta\alpha}^B$ is the Born dynamical effective charge tensor of atom $I$ and $\epsilon_{\beta\gamma}^\infty$ is the high-frequency dielectric tensor. The small-momenta behaviour of the EEL intensity is hence dominated by the $1/q^2$ factor for insulators, since $\boldsymbol{Z}_I(\boldsymbol{q})$ tend to a constant value equal to the longitudinal charges measured in IR spectroscopy[15, 16]. In the opposite large-$q$ limit, the valence screening becomes inefficient and $\boldsymbol{Z}_I(\boldsymbol{q})$ tends to the naked ionic limit $Z_I^{ion}$ in all materials, resulting in similar EEL intensities in metals and insulators alike. At intermediate momenta, the effective charges display an almost linear dependence on $q$ for both h-BN and graphite (see Fig. 2b and Extended Data Fig. 6), that results in an intensity almost independent of $q$, Fig. 2a, irrespectively on the IR polarizability of the target material, at odds with the conventional wisdom[11, 13].

We now show that our large-$q$ phonon EELS technique opens up completely new possibilities. Distinct phonon branches are observed depending on the $q$-path used in the measurements and it is sensitive enough to detect phonon excitations even from a freestanding graphene monolayer. Figure 3a-d present the phonon dispersions measured from a monolayer graphene along the directions ΓMΓM, ΓKMKΓ, Γ→K (the third closest K form the first Γ) and KK (see Fig. 1a) together with the simulated ones (Figs. 3e-h). Although the EEL spectra (Extended Data Fig.7) are relatively noisy compared to the one corresponding to graphite, the intensity colour maps unambiguously draw the phonon dispersion curves of monolayer graphene, which are well consistent with the simulations for all the measured directions. Along the symmetric lines (ΓMΓM (Fig. 3a and 3e) and ΓKMKΓ (Fig. 3a and 3e)), one can access the LO and LA modes analogous to the graphite case. In addition, the TA mode turns detectable at the latter half of the second Brillouin zone along the low symmetric line (Fig. 3c and 3g) due to the symmetry breaking (see also the bubble plots in Extended Data Fig. 8). Such high sensitivity combined with a flexible accessibility to the different modes makes possible the full study of vibrational properties of atomically thin materials. Furthermore, we obtain the dispersion along the KK direction, which cuts across the second Brillouin zone (Fig. 3d and 3h) in expectation of an energy drop of the LO/TO mode at the K point, namely, Kohn anomaly. Interestingly, compared to the theory, our experiment indicates a rather gentle energy drop of LO/TO mode at K points, which entails further detailed analysis that implies better energy and momentum resolutions.

The large-$q$ EELS approach described here has important advantages over the previous reports on momentum integrated STEM-EELS. We are capable of measuring all materials with almost negligible EELS signal delocalization effect at large-$q$ conditions[17], which makes truly localised measurements possible. To further demonstrate the possibilities of local phonon probing for different vibration modes at a specific $q$ we performed measurements on graphene nanoribbons. The beam size in this experiment was approximately 10 nm and the corresponding momentum resolution is ±0.2 Å$^{-1}$. The sample consists of graphene nanoribbons having a few tens nanometre width on a top of 18 ~ 20 graphene layers estimated from the annular dark field (ADF) image profile (Fig. 4). An EEL spectrum obtained at $q$=3.5 Å$^{-1}$ in the direction ΓMΓM is depicted in Figure 4b. The phonon response there consists of two broad peaks: A) one between 80-130 meV related to the LA/ZO mode and B) another for the LO/TO mode at about 180 meV. Very interestingly both signals yield a different response (Fig 4c-e). Whereas the LO/TO EELS intensity simply reflects the specimen thickness and is insensitive on the nanostructure of the sample, the LA/ZO signal is strongly enhanced by the presence of the edges. In addition, structural defects at impurities at surfaces also yield an increased intensity of the lower energy modes (white dashed circles in Fig. 4d). The vibration mode mapping is also possible with a confined single layer graphene as shown in Extended Data Fig. 9. A similar trend of the edge enhancement in the specific vibration mode is also found here. It is noted that the edge vibrations of the nanoribbon and some local sp$^3$ surface vibrations[18] are in the energy range between 80-130 meV. Hence, our experiment represents a proof of principle that demonstrates new possibilities to study the local vibration modes at nanometre-scale on monolayer 2D materials. Spatial and momentum resolved measurements will enable us to fully disentangle different vibration modes and their momentum transfers at non-perfect structures such as edges or defects, which are extremely important to understand materials' local properties. This has been the biggest challenge in electron microscopy since the spatial and momentum resolutions are compensated due to the limit of Heisenbergs incertitude principle. We believe that our methodology wedging this limit will boost vast research in material science.

**Methods**

**STEM-EELS.** The momentum-resolved EELS is performed by using the "parallel-beam scanning mode" in which a parallel electron beam (with a convergence semi-angle $\alpha$ < 0.1 mrad) passes through the sample and makes a crossover at the diffraction plane (Fig. 1a). The entrance aperture of the spectrometer is placed at a given position in the diffraction plane including the first, second or third Brillouin zones and determines the momentum resolution (±0.1~0.2 Å$^{-1}$). The momentum range employed here (0 < $q$ < 5 Å$^{-1}$) is almost three times wider than the conventional inelastic x-ray (neutron) scattering spectroscopy or EELS in reflection, which typically probe the first Brillouin zone only. The experiments were performed by a JEOL TEM (3C2) equipped with a Schottky field emission gun, a double Wien filter monochromator and delta correctors at 30 keV. EEL spectra were collected by STEM mode in which the energy resolution was set to 30 meV in FWHM. In this energy resolution, the peaks at less than 30 meV are hardly discriminated from the quasi elastic line (the grey shadow region in Fig. 1d,e and Fig.3). However this experimentally inaccessible region also depends on the measured $q$. Especially, near the Braggs reflection spots ($\Gamma$ points), the stronger quasi elastic line possibly buries the lower energy peaks. The convergent semi-angle and EELS detection semi-angles are set to 0.2 and 1.1 mrad, respectively. This condition provides the momentum resolution as ±0.1 Å$^{-1}$ and the corresponding probe size as 40 nm. Since the momentum and spatial resolutions are balanced each other, the smaller probe size (~10 nm) as shown in Fig. 4 can be also achievable by integrating the momentum space as ±0.2 Å$^{-1}$. The probe current was 10 pA. The spectra shown in this study for $q$ >0.75 Å$^{-1}$ are constructed by summing 50 spectra in which the dwell time of each spectrum is 30 seconds for graphite and h-BN (Fig.1) and 50 seconds for a single layer graphene (Fig.3).

**Ab-initio calculation.** The effective charges defined in Eq. (2) and phonon dispersions in different materials were calculated within DFPT[10] as implemented in Quantum Espresso[19], using local-density approximation (LDA) and norm-conserving pseudopotentials. A 14×14×6 mesh of k-points and a cutoff of 70 Ry was used for bulk h-BN, while a cutoff of 90 Ry and a mesh 32×32×8 (32×32×1) were used for graphite (graphene) with a Hermite-Gaussian smearing of 0.02 Ry. Phonon dispersion curves have been calculated interpolating the dynamical matrices of 12×12×1 grids, while the effective charges and the EELS intensities have been evaluated for selected $q$-points along given directions in reciprocal space. Notice that, strictly speaking, the quantity defined in Eq. (2) is a complex number, even though the imaginary part is in the present cases much smaller than the real part; indeed, in all calculations we found that the former is always one order of magnitude smaller than the latter (Extended Data Fig.6). The simulated spectra have been calculated using Eq. (1) and smearing the delta function with a Lorenztian broadening of 25 meV to mimic the experimental resolution.

**Acknowledgements:**

This work was supported by KAKENHI (17H04797 and 16H06333) and by the European Graphene Flagship Project. T.P. thanks the FWF P27769-N20 for funding. P.B. and F.M. acknowledge the CINECA awards under the ISCRA initiative (Grants HP10BLTB9A, HP10BSZ6LY), for the availability of high performance computing resources and support. We thank Paola Ayala for fruitful discussions.

**Author contributions**

RS, SM, TP and KS designed the experiments. RS performed EEL spectroscopy. RS and TP analysed data. PB and FM established the theory and performed ab-initio calculations. RS, TP, PB, FM and KS co-wrote the paper. All commented on manuscript.

**Competing interests**

The authors declare no competing financial interests.

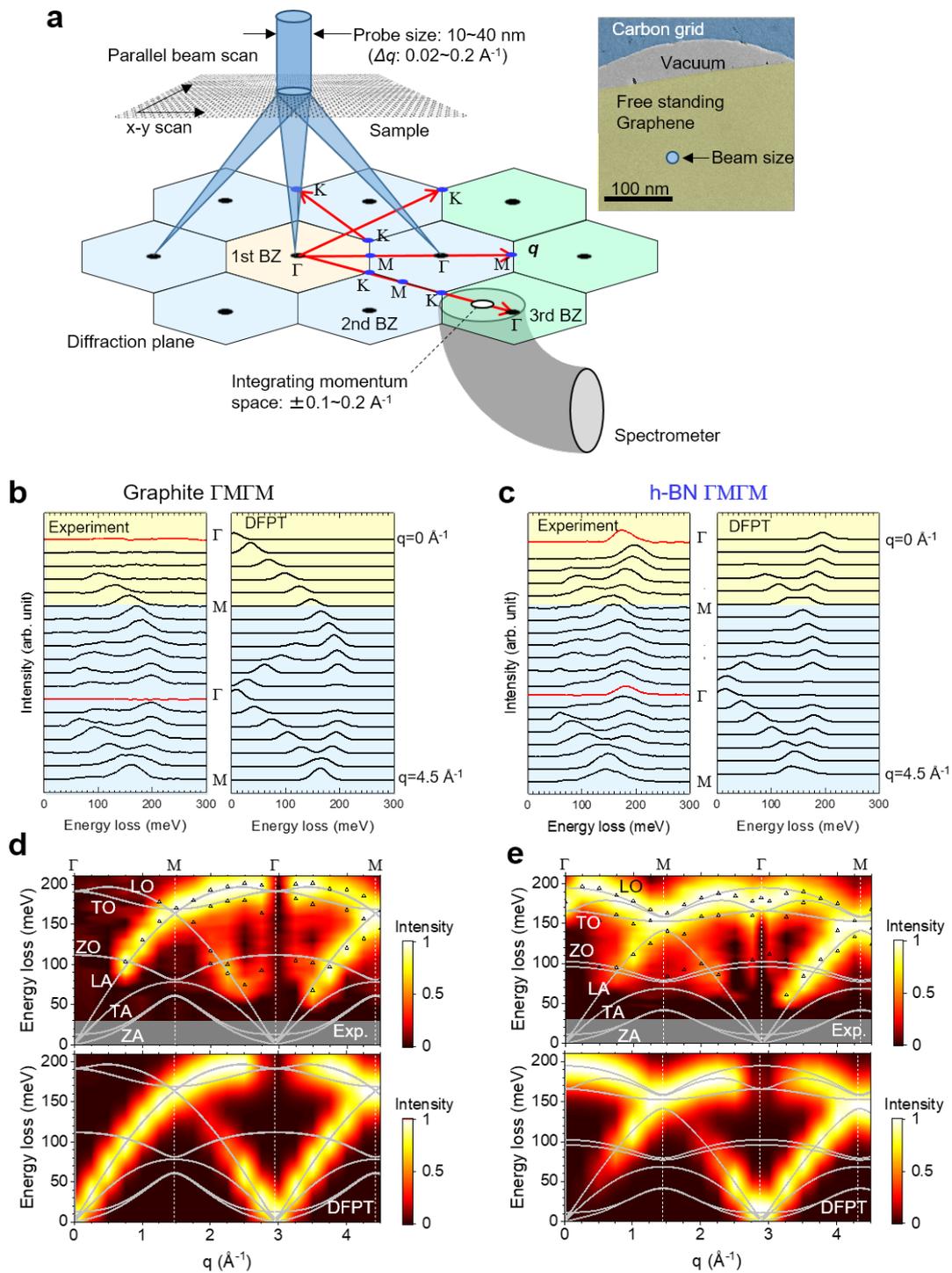

**Figure 1| Momentum resolved vibrational spectroscopy of graphite and h-BN by EELS. a**, Schematic of experimental setup. The EEL spectra are obtained from each spot along several lines with the EELS aperture focused on the diffraction plain where the certain momentum spaces are integrated. **b,c**, Series of momentum resolved EEL spectra

(left) and simulated ones (right) along ΓMΓM direction (from top to bottom) for graphite and h-BN, respectively. The spectra are recorded at every 0.25 Å$^{-1}$ from $q$=0 to 4.50 Å$^{-1}$ in ΓMΓM direction as well as an exception at the second Γ point of h-BN (2.88 Å$^{-1}$). The spectra at every Γ points (red lines) include the Braggs reflection spots. **d,e,** Intensity colour maps of graphite and h-BN constructed from the measured EEL spectra (top) and simulated ones (bottom) are shown with the simulated phonon dispersion curves (solid line). The peak positions extracted from the measured spectra by the line shape analysis are also indicated by the open triangles in the top panels of **d** and **e**. The grey regions in **d** and **e** show experimentally inaccessible energy regions where the peaks are hardly discriminated from the quasi-elastic line.

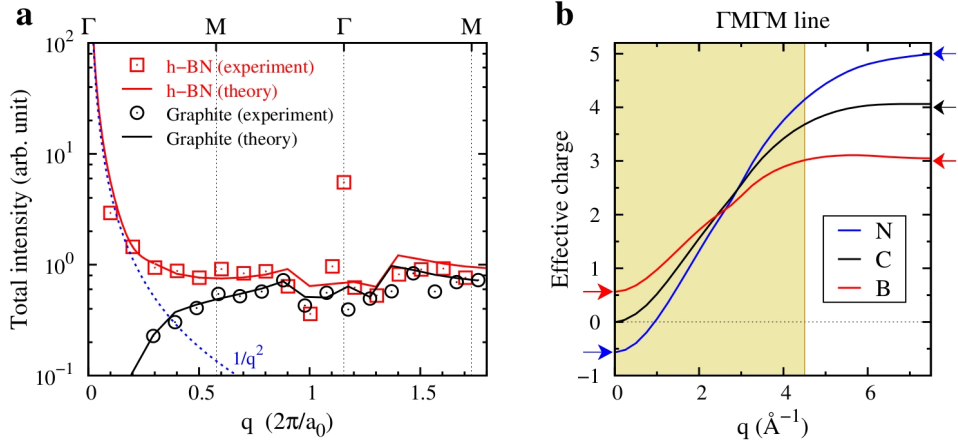

**Figure 2| Total phonon intensity and charge modulation in EELS. a,** The total intensity of graphite (black) and h-BN (blue) as a function of momentum transfer obtained from the experiments (closed circles with broken lines) and DFPT calculations (solid lines), evaluated as $I(\boldsymbol{q}) = \int_{\omega_{cut}}^{\infty} d\omega \frac{d^2\sigma}{d\Omega d\omega}(\boldsymbol{q}, \omega)$, where a lower energy-cutoff $\omega_{cut}$ – corresponding to the grey regions in Fig 1d and 1e – was assumed to ease the comparison. A purple line dropped with $1/q^2$ is shown as a reference. **b,** Evolution of the real part of the component of effective charges parallel to momentum $\boldsymbol{q}$ evaluated along the ΓMΓM line in the polar semiconductor h-BN and in metallic graphite (the small imaginary part is shown in Extended Data Fig. 5). The shaded area indicates the momentum range probed by the experiment, for which the total intensities shown in **a** have been measured. Arrows show the limits expected at small momentum (the longitudinal IR charge given by Eq. 3) and at large momentum (the charge of naked ions), respectively.

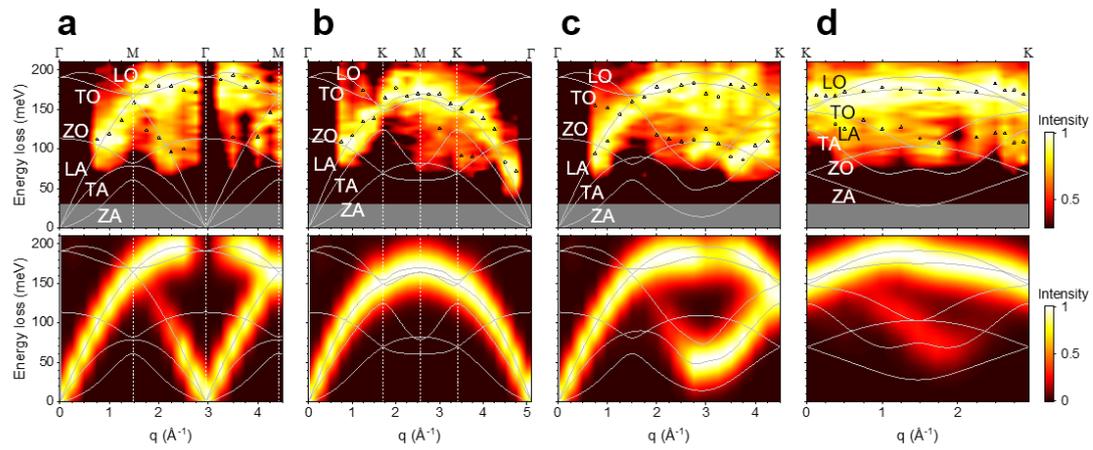

**Figure 3| Momentum mapping of vibrations in graphene monolayer. a-d**, Intensity colour maps constructed from the measured EEL spectra (top) and simulated ones (bottom) together with the phonon energies (open triangles) taken from a single layer graphene along ΓMΓM, ΓKMKΓ, Γ→K (the third closest K form the first Γ) and KK directions (see also sketch in Fig. 1a). The EEL spectra in which the quasi elastic lines are subtracted are shown in Extended Data Fig. 7.

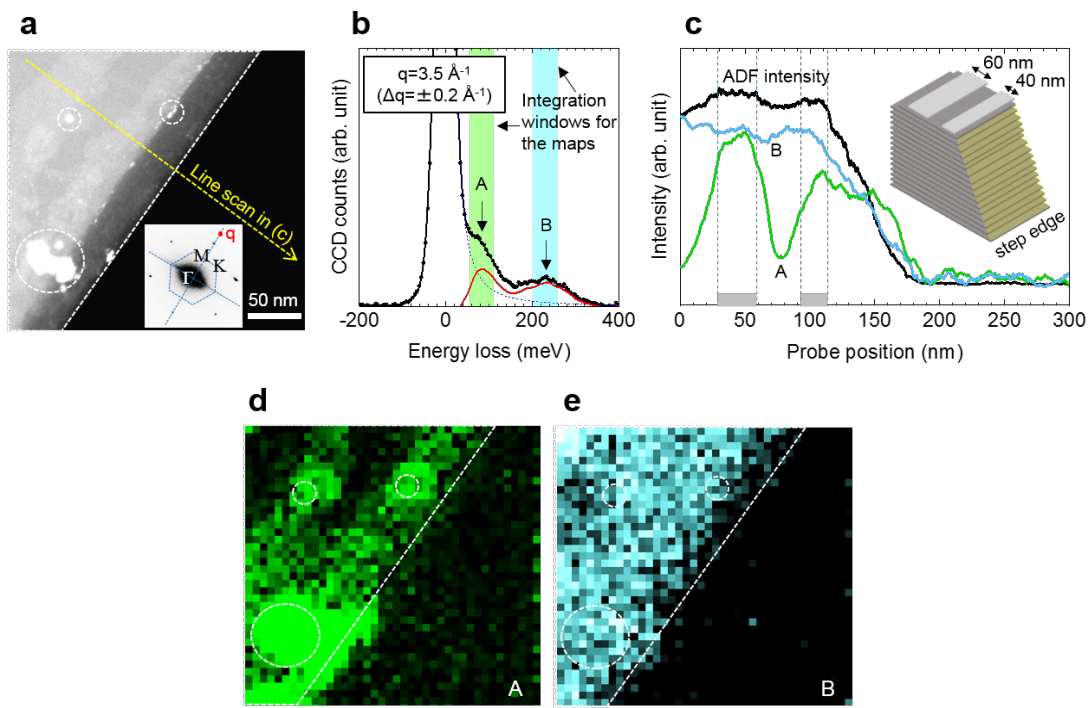

**Figure 4| Position mapping of vibrations in graphene nanoribbons. a**, STEM image of a graphite edge with graphene nanoribbons on it and its diffraction pattern (inset). The guide lines for the edge and the positions of impurities are shown by white broken line in each figure. The red point in the diffraction pattern indicates the place of EELS aperture at $q$=3.5 Å$^{-1}$ where the spectra collected for the position mapping of vibrations. From the diffraction pattern, the steps at the edge should have armchair structures. The measured $q$ is on the parallel direction to the edge. **b,** EELS spectra and integration windows for the lower energy mode (A) including LA/ZO as well as the edge and sp$^3$ defect contributions and the higher energy mode (B) including LO/TO modes are shown in the light green and blue stripes. **c**, Intensity profiles of ADF (black), the lower (red) and higher (blue) energy modes along the yellow line in **a.** The schematic of the edge along this line scan is shown as the inset. **d,e** Corresponding position mapping of the vibration mode A and B, respectively.

Extended data figures

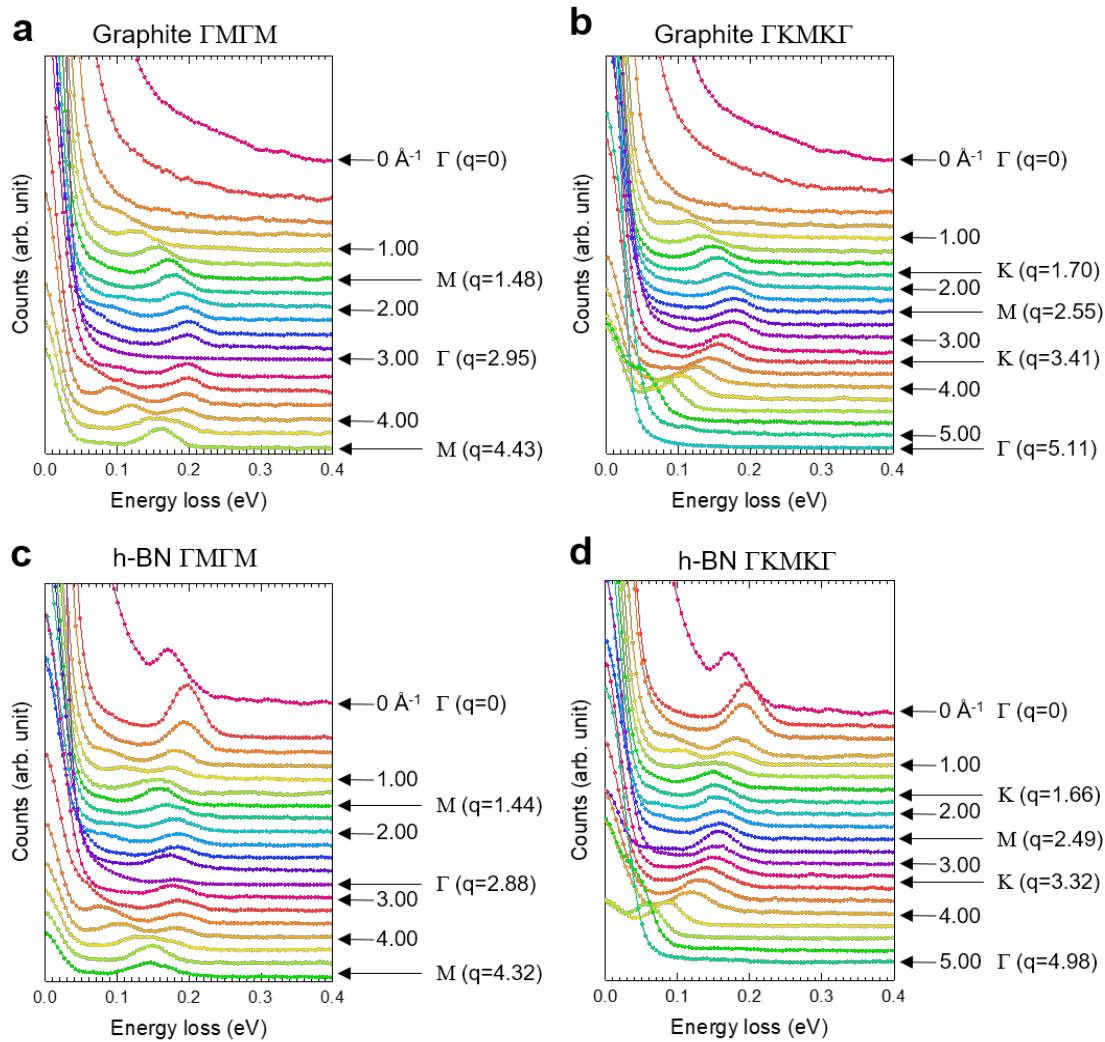

**Extended Data Figure 1| Raw EEL spectra including quasi elastic lines. a–d,** Series of momentum resolved EEL spectra along ΓMΓM (**a**, **c**) and ΓKMKΓ (**b**, **d**) directions taken from the graphite and h-BN flakes.

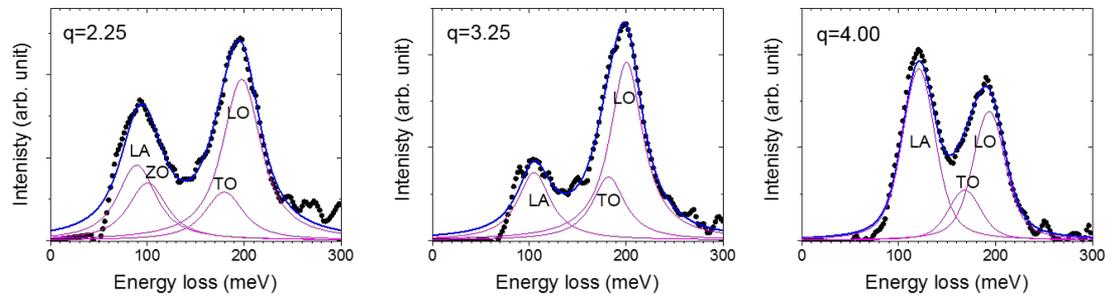

**Extended Data Figure 2| Line shape analysis on the measured spectra.** Typical examples for the line shape analysis on the EEL spectra at $q$=2.25, 3.25 and 4.00. The each component (the purple lines) is fitted with Voightians including the fixed experimental broadening factor (30~40 meV).

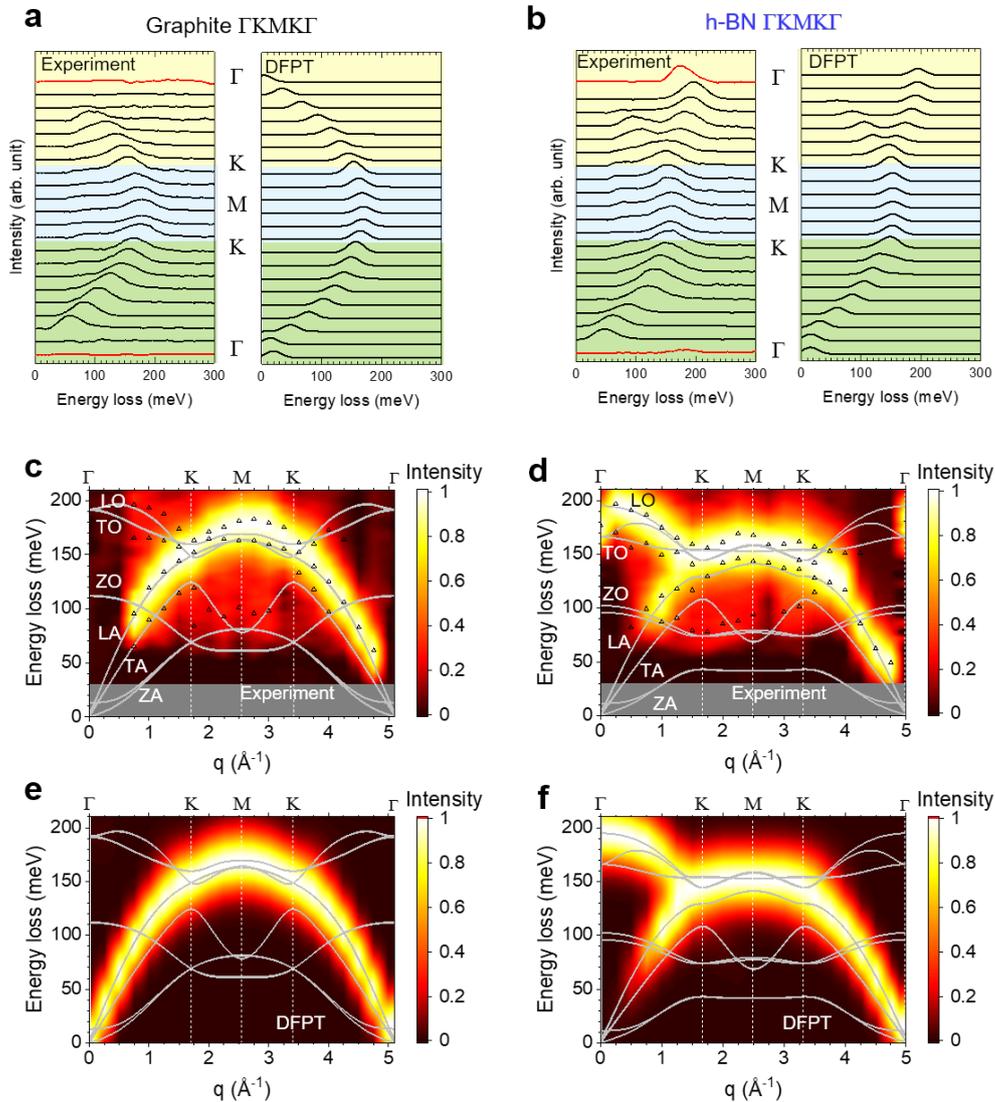

**Extended Data Figure 3| Phonon dispersion of h-BN and graphite along ΓKMKΓ line. a,b,** Series of momentum resolved EEL spectra (left) and simulated ones (right) along ΓKMKΓ direction (from top to bottom) for graphite and h-BN, respectively. The quasi elastic lines are subtracted (The raw data including the quasi elastic lines are shown in Extended Data Fig. 1). The spectra are recorded at every 0.25 Å$^{-1}$ from $q$=0 to 5.00 Å$^{-1}$ in ΓM direction as well as an exception at the second Γ point of graphite (5.11 Å$^{-1}$). The spectra at every Γ points (red lines) include the Braggs reflection spots. **c-f,** The intensity colour maps of the graphite and h-BN constructed from the measured EEL spectra (**c,d**) and the simulated ones (**e,f**) are shown with the simulated phonon dispersion curves (solid lines). The peak positions extracted from the measured spectra by the line shape analysis are indicated by the open triangles in **c** and **d**.

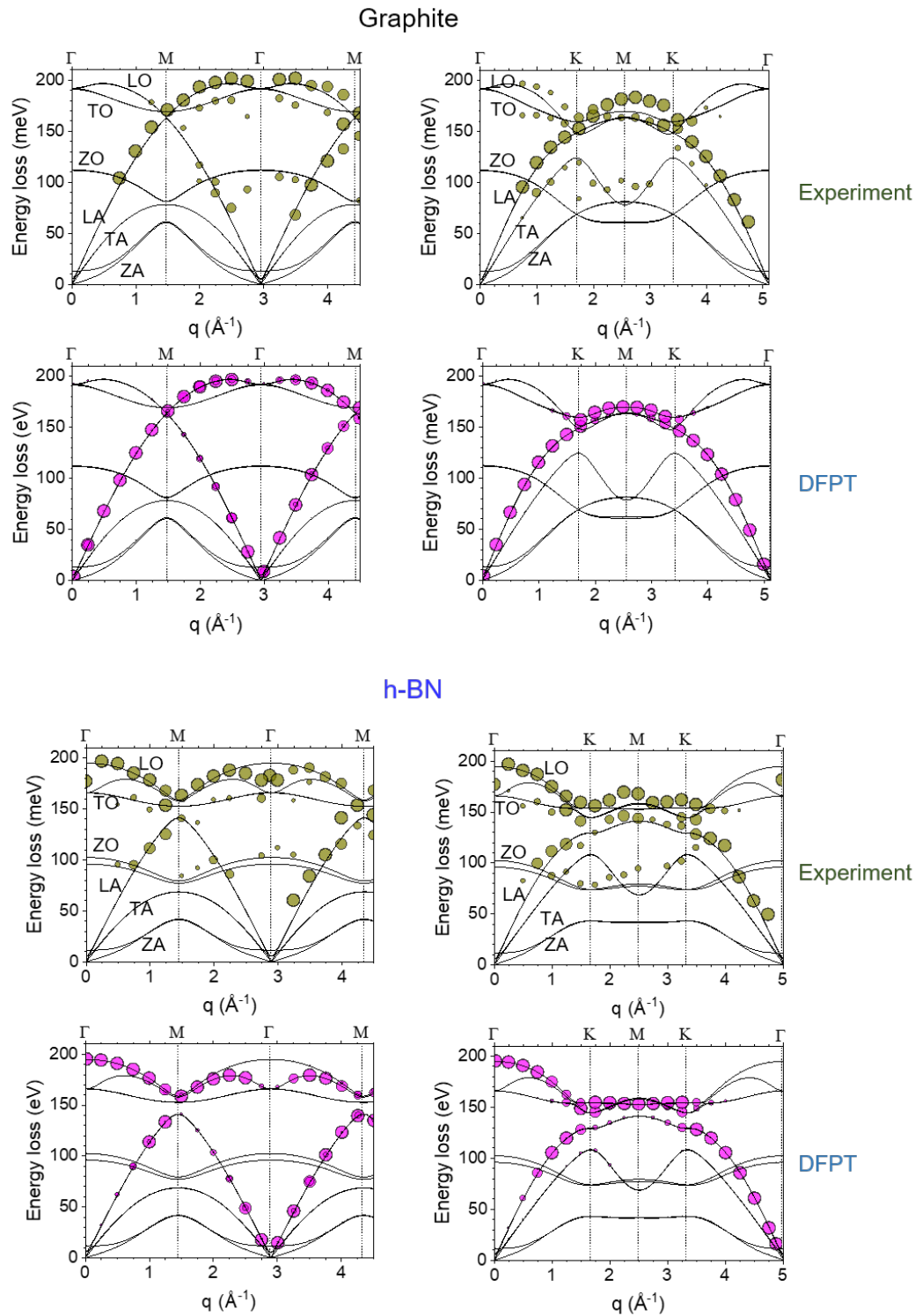

**Extended Data Figure 4| Bubble plots for the phonon dispersion of graphite and h-BN.** These Plots obtained by the line shape analysis on the measured spectra (upper panels) and from the simulation (bottom panels). The plot size (cross section) corresponds to the intensity which is normalized by the highest peak at each measured (calculated) $q$.

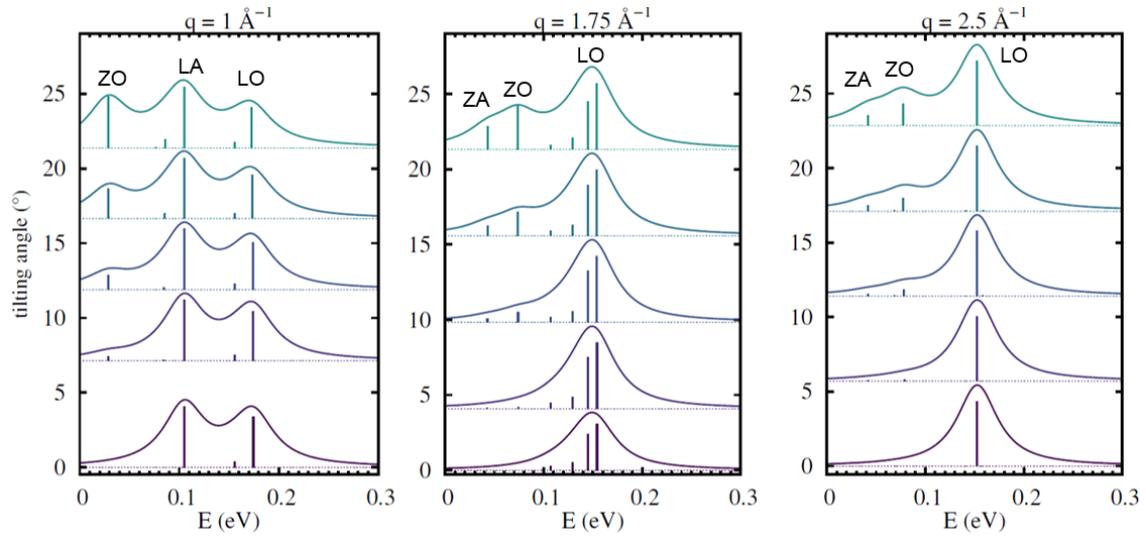

**Extended Data Figure 5| The contribution of out of plane mode by the sample tilting.** The simulated spectra of h-BN at $q$=1, 1.75 2.5 Å$^{-1}$ along the ΓKMKΓ line considering the sample tilt. ZA and ZO modes are gradually activated as the tilting angle increases. Bars indicate the mode-resolved calculated intensities at each phonon frequency, that are then broadened with a 25 meV Lorentzian resulting in the simulated spectra.

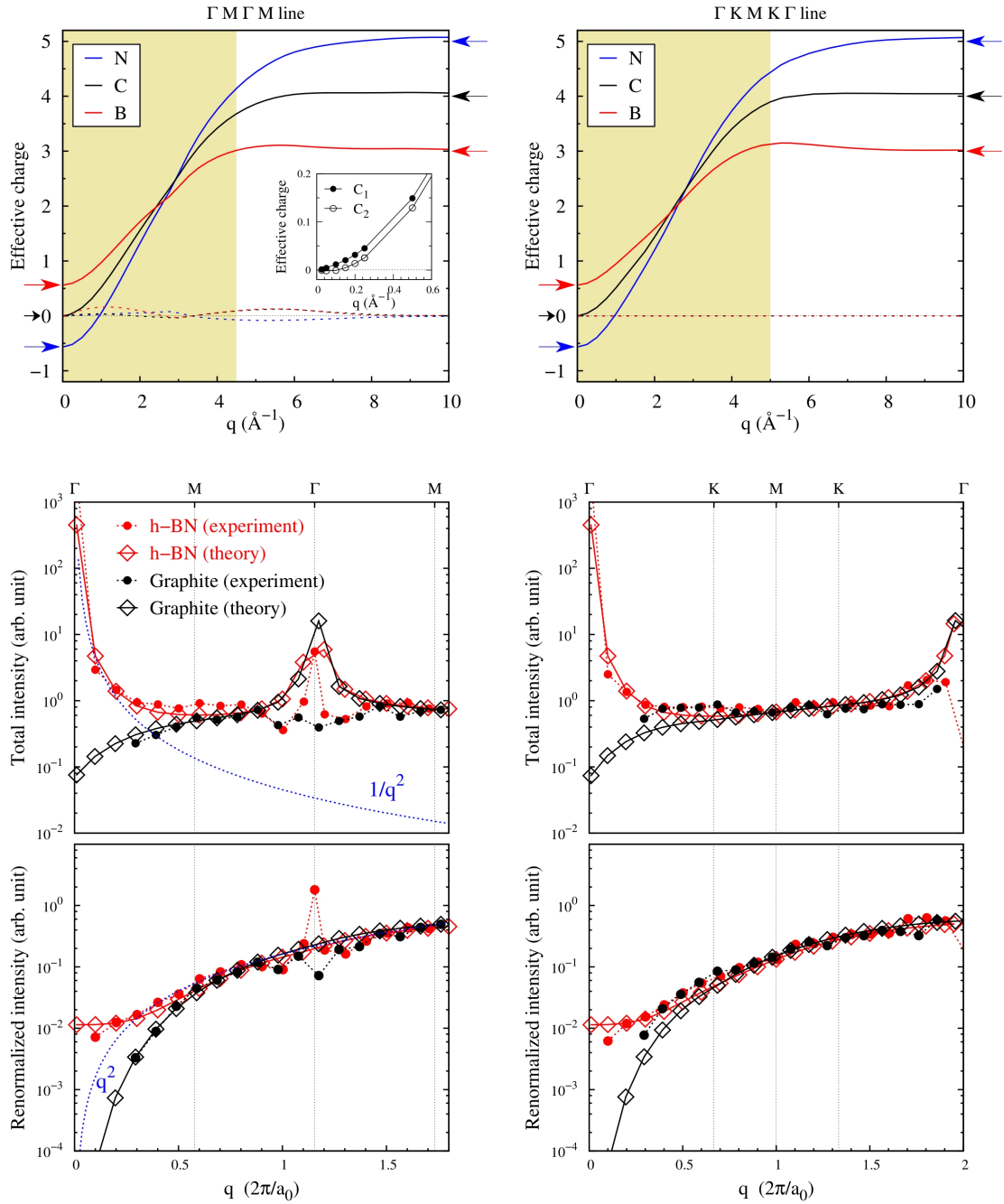

**Extended Data Figure 6| *q*-dependence of effective charges and total intensity for phonon EEL spectra. Top,** The evolution of effective charges in both h-BN and graphite as a function of *q* are shown along the $\Gamma M \Gamma M$ and $\Gamma K M K \Gamma$ directions. Solid (dashed) lines correspond to the real (imaginary) part of the effective charge. Despite graphite hosts two inequivalent Carbon ions, their effective charges are substantially undistinguishable but for very small momenta, where only tiny differences can be seen, as shown in the inset. **Bottom,** The total intensity for graphite (black) and h-BN (blue) as a function of momentum transfer obtained from the experiments (closed circles with broken lines) and

DFPT calculations (solid lines), evaluated as $I(\boldsymbol{q}) = \int d\omega \frac{d^2\sigma}{d\Omega d\omega}(\boldsymbol{q}, \omega)$. The singular behaviour of the theoretical total intensity observed at G points in higher-order Brillouin zones can be ascribed to the $1/\omega_{qv}$ dependence of the differential cross section, which diverges for LA modes (notice that this diverging behaviour does not appear if the total intensity is evaluated with a lower-energy cutoff $\omega_{cut}$, as shown in Fig 2a). A renormalized intensity defined as $I(\boldsymbol{q}) \propto \sum_v \left| \sum_l \frac{1}{\sqrt{M_l}} \boldsymbol{Z}_l(\boldsymbol{q}) \cdot \boldsymbol{e}_{\boldsymbol{q},v}^l e^{-i\boldsymbol{q}\cdot\boldsymbol{\tau}_l} \right|^2$ is also shown, which allows to highlight the non-trivial momentum-dependence of EEL intensity stemming from effective charges by neglecting the trivial momentum and phonon dependencies $1/q^2$ and $(1+n_{qv})/\omega_{qv}$, respectively. A purple line scaling as $1/q^2$ ($q^2$) is also shown as a reference for total (renormalized) intensities.

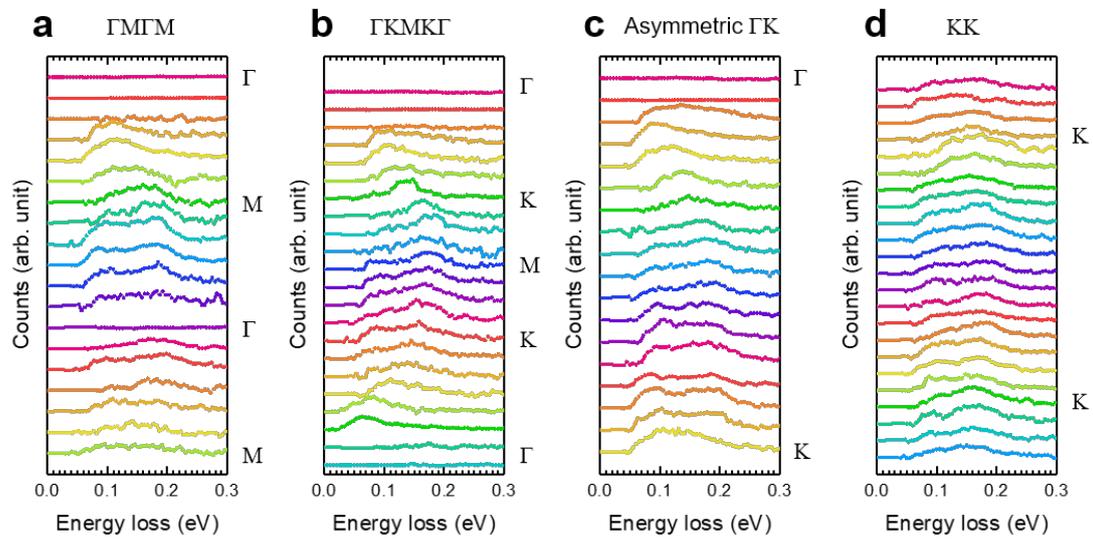

**Extended Data Figure 7| Phonon EEL spectra of a single-layer graphene. a-d**, Series of momentum resolved EEL spectra along (**a**) ΓMΓM, (**b**) ΓKMKΓ, (**c**) asymmetric ΓK and (**d**) KK directions taken from a single-layer graphene. The quasi elastic lines are subtracted.

## Graphene ΓΜΓΜ

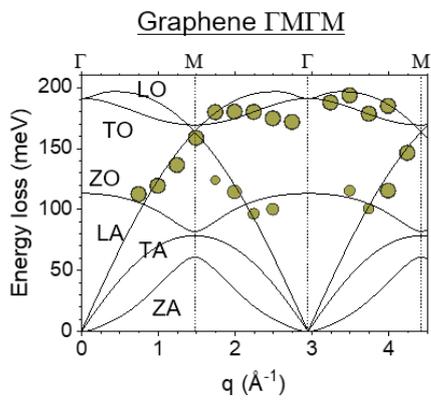

## Graphene ΓΚΜΚΓ

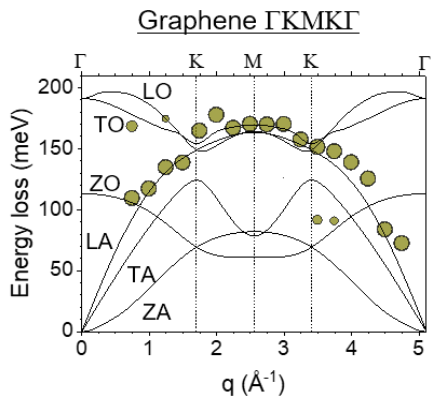

Experiment

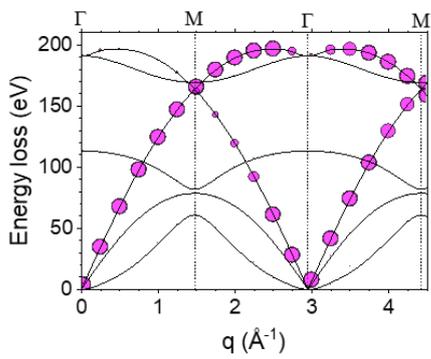

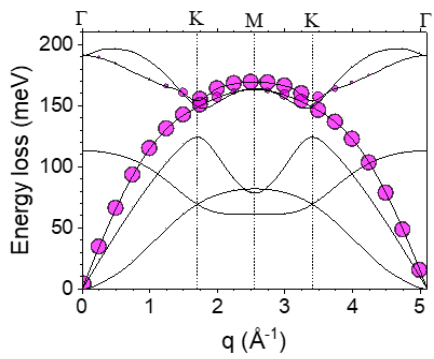

DFPT

## Graphene asymmetric ΓΚ

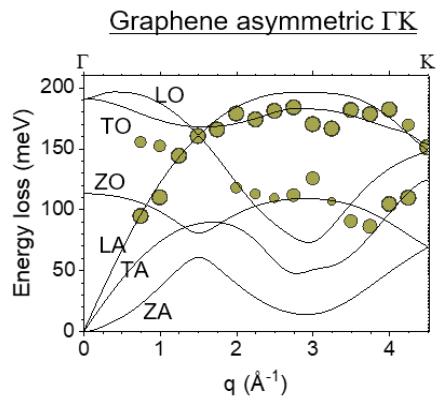

## Graphene ΚΚ

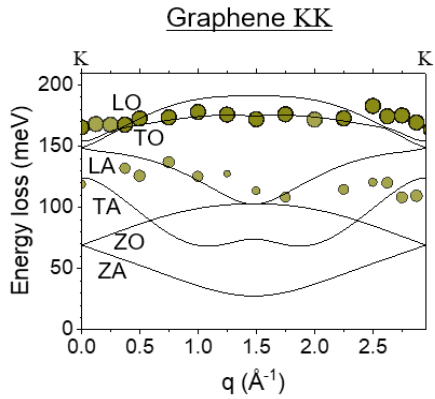

Experiment

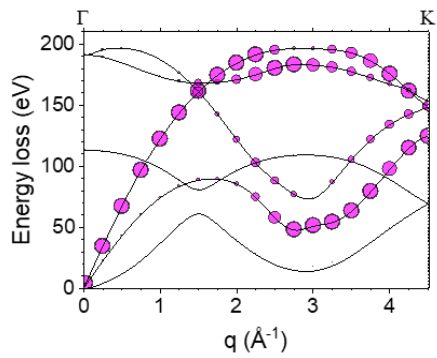

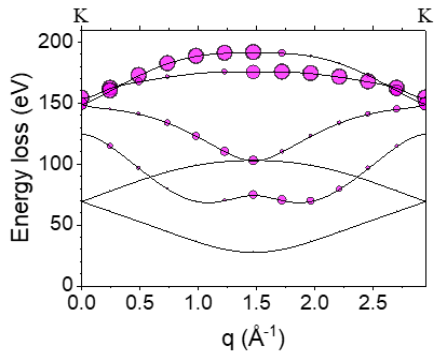

DFPT

**Extended Data Figure 8| Bubble plots for the phonon dispersion of graphene monolayer.** The top panels exhibit the plots obtained by the line shape analysis on the measured spectra. Whereas the bottom panels present the plots obtained by the simulation. The measured (simulated) directions are ΓMΓM (**top left**), ΓKMKΓ (**top right**), Γ → the third closest K (**bottom left**) and KK (**bottom right**). The plot size (cross section) corresponds to the intensity which is normalized by the highest peak at each measured (calculated) $q$. In ΓMΓM line, LO modes (170-200 meV) are visible at the second Brillouin zone besides LA modes which shows up throughout. In ΓKMKΓ direction, TO mode is additionally seen at the second Brillouin zone along KMK line for both experiment and simulation. In the low symmetric line (Γ → the third closest K), TA mode shows up in the latter half of the second Brillouin zone. The dispersion along KK direction which cuts across the second Brillouin zone show LO mode in the first half and TO mode in the later half in the simulation while this asymmetric behaviour is not clearly seen in the experiment. LA and TA modes are also activated weakly at the middle of KK line both in experiment and simulation.

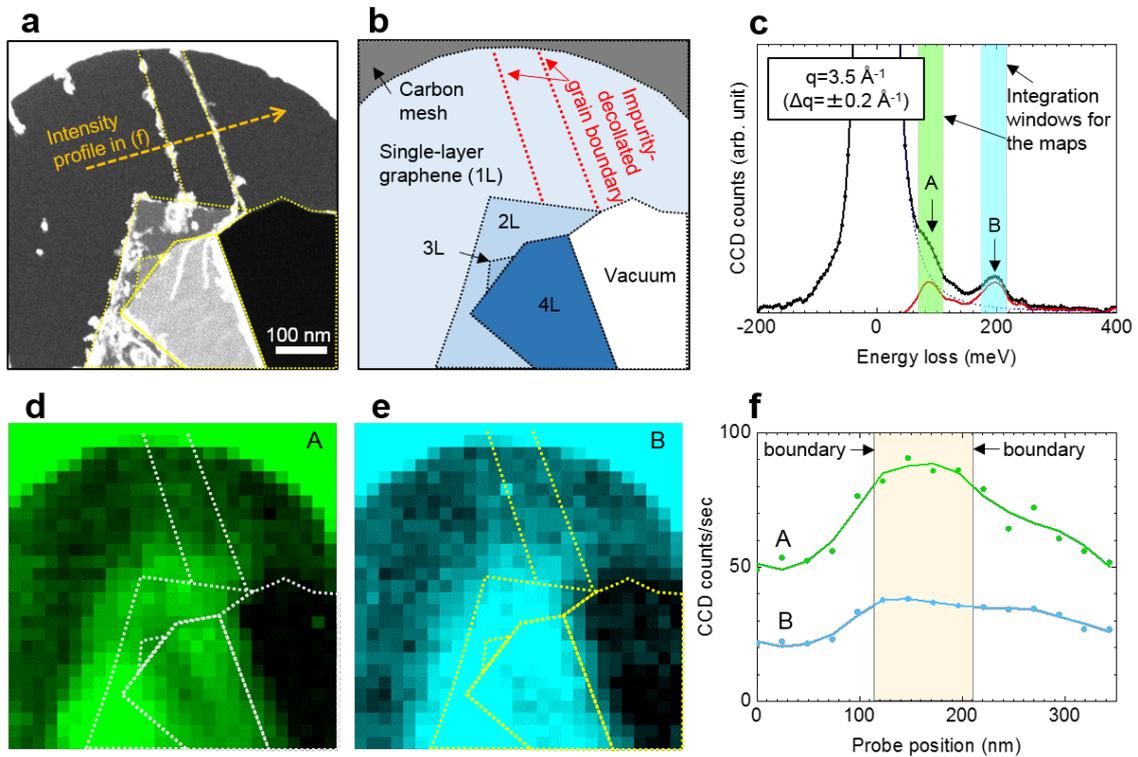

**Extended Data Figure 9| Position mapping of vibrations in graphene monolayer. a,
b** STEM image and its schematic for a free-standing graphene having double-, triple and
four-layer regions as well as a single-layer region. **c**, Typical EEL spectrum at $q$=3.5 Å$^{-1}$
taken from a clean single-layer region where 16 pixels are integrated. The phonon
response there consists of two broad peaks: A) one between 80-130 meV related to
LA/ZO mode as well as the edge and sp$^3$ defect contributions and B) the other for LO/TO
mode at about 180 meV. **d, e**, Corresponding position mappings of vibration modes A and
B, respectively. The integration windows for the mappings are also shown in the light
green and blue stripes in **c. f**, Intensity profiles of vibration mode A and B along the orange
line in **a** across the two impurity decollated grain boundaries. LO/TO mode intensity (B)
is simply proportional to the layer number (the specimen thickness), while the lower
energy mode (A) clearly shows the higher intensity at the edges of double- and four-layer
regions. Interestingly no obvious intensity enhancement can be seen at the edge of single-
layer. In addition we found the intensity variations of the low energy mode at grain
boundaries as shown in **f**, though the interpretation is more complicated since there must
be contributions of different momentum transfers derived from the domains with different
orientations.